\documentstyle[preprint,revtex]{aps}                  

\newcommand{\be}{\begin{equation}}
\newcommand{\ee}{\end{equation}}
\newcommand{\beq}{\begin{eqnarray}}
\newcommand{\eeq}{\end{eqnarray}}

\begin{document}
\draft
\pagestyle{empty}                                      
\centerline{
                             \hfill   NTUTH--95--04}   
\centerline{\hfill                 May 1995} 
\vfill
\begin{title}
Magnetic Field Dependence of Muonium-antimuonium Conversion
\end{title}
\vfill
\author{Wei-Shu Hou and Gwo-Guang Wong}
\begin{instit}
Department of Physics, National Taiwan University,
Taipei, Taiwan 10764, R.O.C.
\end{instit}
%
\vfill
\begin{abstract}

We study the magnetic field dependence of
muonium--antimuonium conversion induced by
neutral (pseudo)scalar bosons.
Only the $SS$ operator contributes to
the conversion of polarized muonium,
but it gets quenched by a magnetic field
of strength 0.1 Gauss or stronger.
Conversion induced by $SS$ couplings for unpolarized muonium
is independent of magnetic field.
Magnetic fields of 0.1 Tesla or stronger
starts to suppress conversion induced by
$PP$ interactions in the lowest Breit-Rabi level,
but gets partially compensated by a rise in conversion probability
in the other unpolarized level.
The effects of $(S\mp P)(S\mp P)$ and $(S\mp P)(S\pm P)$
operators behave in the same way as
$(V\mp A)(V\mp A)$ and $(V\mp A)(V\pm A)$ operators,
respectively.

\end{abstract}
\newpage
\narrowtext
\pagestyle{plain}


The spontaneous conversion of muonium
(hydrogen-like atom $M = \mu^+e^-$)
into antimuonium ($\bar M$)
would violate the separate additive muon
and electron numbers,
but would remain consistent with
multiplicative muon or electron number conservation.
Defining the effective coupling $G_{M\bar M}$  via
the interaction \cite{FeinWein},
\begin{equation}
{\cal H}_{M\bar M} = {G_{M\bar M} \over \sqrt{2}}
                                         \bar \mu \gamma_\lambda (1-\gamma_5) e
\,
                                         \bar \mu \gamma^\lambda (1-\gamma_5) e
+ \mbox{h.c.},
\end{equation}
the limit has just been improved \cite{Jungmann}
by an order of magnitude,
\begin{equation}
G_{M\bar M} < 1.8\times 10^{-2}\ G_F,
\end{equation}
compared with the previous bound of $0.16\ G_F$ \cite{Matthias1},
where $G_F$ is the Fermi constant.
The ultimate aim \cite{PSI} is to
reach the sensitivity level of $10^{-3}\ G_F$.

The $(V-A)(V-A)$ interaction of eq. (1)
gained further theoretical footing when
Halprin \cite{Halprin,H++} pointed out that
in left-right symmetric models with Higgs triplets,
doubly charged scalars can mediate
$M$--$\bar M$ transitions at tree level.
The effective interaction is of $(V\pm A)(V\pm A)$ form
after Fierz rearrangement.
The possibility of
$(V-A)(V+A)$ interactions,
induced by dilepton gauge bosons \cite{dilepton},
was discussed by Fujii {\it et al.} \cite {Fujii}.
Interestingly, muonium conversion is
more pronounced in the singlet channel,
in contrast to the $(V-A)(V-A)$ case where
the conversion matrix element is of equal strength
for both singlet and triplet muonium \cite{FeinWein}.

Recently,
Horikawa and Sasaki \cite{Sasaki}
pointed out further that
$(V-A)(V+A)$ interactions are 
considerably less senstive
to magnetic fields compared to the $(V-A)(V-A)$ case.
Since actual experiments are conducted
in the presence of magnetic fields \cite{Jungmann,Matthias1,PSI},
this has important implications
for the interpretation of
the bound of eq. (2) for $(V-A)(V+A)$ interactions.
In a recent paper \cite{HW} (see also \cite{Derman,Matthias2}),
we explored neutral (pseudo)scalar induced
$M$--$\bar M$ oscillations.
Since $(S\mp P)(S\pm P)$ and
$(V\mp A)(V\pm A)$ operators are related
by Fierz transform,
we wish to study the magnetic field dependence of
neutral scalar induced muonium conversion.

Exotic neutral (pseudo)scalar bosons $H$ and $A$
could have couplings \cite{HW},
\begin{equation}
-{\cal L}_{Y} = {f_H \over \sqrt{2}}\, \bar \mu e\, H
                     + i {f_A \over \sqrt{2}}\, \bar \mu \gamma_5 e \, A +
h.c.,
\end{equation}
while a discrete symmetry such as multiplicative \cite{FeinWein}
electron number $P_e$ \cite{CK}
forbids processes odd in number of electrons (plus positrons)
like $\mu\to e\gamma$ and $\mu\to ee\bar e$.
The resulting effective Hamiltonian responsible for
$M$--$\bar M$ conversion is,
\begin{eqnarray}
{\cal H}_{M\bar M} &=& {f_H^2 \over 2m_H^2}\, \bar \mu e\, \bar \mu e
                           - {f_A^2 \over 2m_A^2}\, \bar \mu \gamma_5 e\, \bar
\mu \gamma_5 e \cr
                  &=& {1\over 4} \left({f_H^2 \over m_H^2} + {f_A^2 \over
m_A^2}\right)
                                                  (S^2-P^2)
                         + {1\over 4} \left({f_H^2 \over m_H^2} - {f_A^2 \over
m_A^2}\right)
                                                  (S^2+P^2),
\end{eqnarray}
where $S = \bar \mu e$ and $P = \bar \mu \gamma_5 e$.
Note that $S^2 - P^2$ and $S^2 + P^2$ contain
$(S\mp P)(S\pm P)$ and $(S\mp P)(S\mp P)$ interactions,
respectively.
In the ``$U(1)$ limit"
of $f_H = f_A$ and $m_H = m_A$ the
subleading $(S\mp P)(S\mp P)$ terms
are completely absent.

The matrix elements of eq. (4) and the accompanying phenomenology
of eq. (3) have been discussed in ref. \cite{HW}.
To explore magnetic field dependence of muonium conversion
probabilities, consider the Hamiltonian for $1S$ muonium,
\begin{equation}
{\cal H} = {\cal H}_0 + a\, \mbox{\boldmath $s$}_e\cdot
                                               \mbox{\boldmath $s$}_{\bar\mu}
                                      -        \mbox{\boldmath $\mu$}_e\cdot
                                               \mbox{\boldmath $B$}
                                      -        \mbox{\boldmath
$\mu$}_{\bar\mu}\cdot
                                               \mbox{\boldmath $B$}
                                      + {\cal H}_{M\bar M},
\end{equation}
where $ {\cal H}_0$ gives the $1S$ energy
$E_0 = -\alpha^2 m/2$, with reduced mass $1/m = 1/m_e + 1/m_\mu$,
$a \cong 1.846\times 10^{-5}$ eV is the
$1S$ muonium hyperfine splitting,
and
$\mbox{\boldmath $\mu$}_e = - g_e \mu_B\, \mbox{\boldmath $s$}_e$, 
$\mbox{\boldmath $\mu$}_{\bar\mu} =
                                                    + g_\mu \mu_B {m_e\over
m_\mu}\,
\mbox{\boldmath $s$}_\mu$,
where $g_e \cong g_\mu \cong 2$ and
$\mu_B = e/(2m_e) \cong 5.788\times 10^{-9}$ eV/Gauss
is the Bohr magneton.
Introducing the dimensionless parameters,
\begin{equation}
X,\ Y \equiv {\mu_B B\over a}\left(g_e \pm {m_e\over m_\mu}g_\mu\right),
\end{equation}
we see that $\vert Y\vert$ is just $1\%$ smaller than $\vert X\vert$.
Ignoring ${\cal H}_{M\bar M}$ for the moment,
the four muonium Breit-Rabi energy levels \cite{Hughes} are,
\begin{eqnarray}
E_M(1,\pm1) &=& E_0 + {a\over 2}\left(\ {1\over 2} \pm Y\right), \cr
E_M(1,\ \ 0) &=& E_0 + {a\over 2}\left(-{1\over 2}  + \sqrt{1+X^2}\right), \cr
E_M(0,\,\ \ 0) &=& E_0 + {a\over 2}\left(-{1\over 2}   - \sqrt{1+X^2}\right),
\end{eqnarray}
which correspond to the eigenstates
\begin{eqnarray}
\vert M;1,\pm1\rangle &=& \,\ \ \ \ \vert M;\uparrow\uparrow \rangle,
                                                         \ \vert M;
\downarrow\downarrow \rangle, \cr
\vert M;1,\,\ \ 0\rangle &=& \,\ \ c\,\vert M;\uparrow\downarrow \rangle
                                            + s\,\vert M;\downarrow\uparrow
\rangle, \cr
\vert M;0,\,\ \ 0\rangle &=& - s\,\vert M;\uparrow\downarrow \rangle
                                             + c\,\vert M;\downarrow\uparrow
\rangle,
\end{eqnarray}
where the magnetic field dependent ``rotation" is
\begin{equation}
s    
   = {1\over \sqrt{2}}\left[1 - {X\over \sqrt{1+X^2}}\right]^{1\over 2},\ \ \
c    
   = {1\over \sqrt{2}}\left[1 +{X\over \sqrt{1+X^2}}\right]^{1\over 2}.
\end{equation}
We have labeled the Breit-Rabi energy levels
with the weak field basis $\vert M;F,m_F \rangle$,
i.e. the corresponding zero $B$ field ($X$, $Y \longrightarrow 0$ and
$s$, $c \longrightarrow 1/\sqrt{2}$)
hyperfine states.
The ``uncoupled" basis of $\vert M;m_{s_e},m_{s_{\bar\mu}} \rangle$
corresponds to the strong field limit of $X,\ Y \gg 1$  and
$s \longrightarrow 0$.
In this limit,
$\vert M;1,0\rangle \longrightarrow \vert M;\uparrow\downarrow \rangle$ and
$\vert M;0,0\rangle \longrightarrow \vert M;\downarrow\uparrow \rangle$,
and the electron and muon hyperfine spin-spin coupling
is overwhelmed by the Zeeman effect.

For antimuonium,
again ignoring the effect of ${\cal H}_{M\bar M}$,
retaining the spin labels and with uncoupled basis
$\vert \bar M;m_{s_{\bar e}},m_{s_\mu} \rangle$,
one simply flips $X\to -X$, $Y\to -Y$ and
hence interchange $s\leftrightarrow c$
in eqs. (8) and (9).
The notable changes are
\begin{equation}
E_{\bar M}(1,\mp1) = E_M(1,\pm1),
\end{equation}
since for given spin, the antiparticle magnetic moments have flipped sign,
and,
\begin{eqnarray}
\vert \bar M;1,0\rangle &=&
                              \,\ \ s\,\vert \bar M;\uparrow\downarrow \rangle
                                  + c\,\vert \bar M;\downarrow\uparrow \rangle,
\cr
\vert \bar M;0,0\rangle &=&
                                   - c\,\vert \bar M;\uparrow\downarrow \rangle
                                  + s\,\vert \bar M;\downarrow\uparrow \rangle.
\end{eqnarray}
We have the following energy differences between
$M$ and $\bar M$ eigenstates,
\begin{eqnarray}
E_M(1,\pm1) - E_{\bar M}(1,\pm1) &=& \pm a\, Y, \cr
E_M(1,\,\ \ 0) - E_{\bar M}(1,\,\ \ 0)     &=&
                     \ \ \ E_M(0,\,\ \ 0) - E_{\bar M}(0,\,\ \ 0) \ \,= 0, \cr
E_M(1,\,\ \ 0) - E_{\bar M}(0,\,\ \ 0)  &=& -(E_M(0,\,\ \ 0) - E_{\bar M}(1,\ \
0))
                                                                  = a\,
\sqrt{1+X^2}.
\end{eqnarray}

The effect of ${\cal H}_{M\bar M}$
can be treated as a perturbation.
Define generically
\begin{equation}
\langle \bar M \vert
                        {\cal H}_{M\bar M} \vert M \rangle = \delta/2
\end{equation}
(for simplicity, we take $\delta$ to be real)
between any two Breit-Rabi $M$ and $\bar M$ energy eigenstates,
it was shown by Feinberg and Weinberg \cite{FeinWein} that
the time integrated probability for an initial muonium state
to decay as antimuonium is
\begin{equation}
P(\bar M) = {\delta^2 \over 2(\delta^2 + \Delta^2 + \lambda^2)},
\end{equation}
where
$\Delta = E_M - E_{\bar M}$
is the energy difference, and $\lambda \cong 2.996\times 10^{-10}$ eV
is the muon decay rate.
The physics is clear: muonium oscillation has to compete with
muon decay and the damping from oscillations
between two states that are very disparate in energy.
The total transition probability is
\begin{equation}
P_T(\bar M) =  \sum_{F,m_F}\vert c_{F,m_F}\vert^2\, P^{(F,m_F)}(\bar M)
\end{equation}
where $\vert c_{F,m_F}\vert^2$ are the populations
in muonium states of eq. (8),
and $P^{(F,m_F)}(\bar M)$ are the probabilities
for an initial $(F,m_F)$ muonium state to decay as antimuonium.

In principle $P^{(1,0)}(\bar M)$ and $P^{(0,0)}(\bar M)$
also contain the probabilities of initial $(1,0)$ or $(0,0)$
muonium states to decay as $(0,0)$ or $(1,0)$ antimuonium,
respectively, since $(1,0)$ and $(0,0)$ are mixtures
of unpolarized states.
To show that these are vanishingly small,
it is useful to notice the hierarchy
\begin{equation}
\delta \ll \lambda \ll a,
\end{equation}
the first of which follows from eq. (2)
for any $ {\cal H}_{M\bar M}$ model.
Combining eqs. (12) and (14) we
see that
$(1,0)\to(0,0)$ and $(0,0)\to (1,0)$ transitions are
extremely suppressed by hyperfine splitting
even for $B = 0$,
and need not be considered.
Similarly, although $(1,\pm 1) \to (1,\pm 1)$ transitions
contribute in $B = 0$ limit,
they become rapidly suppressed even for
rather weak magnetic fields \cite{FeinWein},
because of the mismatch between Zeeman energy levels
for $M$ vs. $\bar M$ polarized states with same $m_F$.

In this work we will discuss only the {\it relative} magnetic field
dependence of $M$-$\bar M$ conversion probabilities.
Differences in coupling strength for various effective operators
at zero $B$ field can be found in refs. \cite{Halprin,H++,Fujii,HW}.
Hence,
{\it we normalize effective interactions
to the conversion probability due to
eq. (1)
at zero magnetic field},
and in particular for equally populated (1/4 each) Breit-Rabi states
(the latter condition would be removed at the end).
Thus, the zero field total transition probability is
\begin{equation}
P_T(\bar M;\ B=0) \simeq {\bar \delta^2 \over 2\lambda^2}
              =      2.56\times 10^{-5} \left({G_{M\bar M}\over G_F}\right)^2,
\end{equation}
which defines the parameter $\bar \delta$.

For the $(V-A)(V-A)$ case one basically multiplies
the r.h.s. (right hand side) of eq. (13) by
$\delta_{m_{s_{\bar e}} m_{s_e}} \delta_{m_{s_\mu} m_{s_{\bar \mu}}}$,
hence
\begin{eqnarray}
\langle\bar M;1,\pm1\vert  {\cal H}_{M\bar M} \vert M;1,\pm1\rangle
                              &=& {\bar\delta\over 2}, \cr
\langle\bar M;1,0\vert  {\cal H}_{M\bar M} \vert M;1,0\rangle =
\langle\bar M;0,0\vert  {\cal H}_{M\bar M} \vert M;0,0\rangle
                              &=& {\bar\delta\over 2\sqrt{1+X^2}},
\end{eqnarray}
and all other matrix elements vanish.
One therefore finds the result \cite{Schaefer}
\begin{eqnarray}
P^{(1,\pm1)}(\bar M) &=&
                    {\bar\delta^2 \over 2(\bar\delta^2 + a^2Y^2 + \lambda^2)},
\cr
P^{(1,\ \ 0)}(\bar M)      = P^{(0,\ \ 0)}(\bar M) &=&
                    {\bar\delta^2 \over 2(\bar\delta^2  + \lambda^2\,
(1+X^2))}.
\end{eqnarray}
The magnetic field dependence for
$P_T(\bar M)$, as well as the separate probabilities
$P^{(1,\pm1)}(\bar M)$, $P^{(1,0)}(\bar M)$ and $P^{(0,0)}(\bar M)$
of eq. (19), are plotted
as ``{\scriptsize $+$}" symbols in Figs. 1--4, respectively.
The behavior is readily understood.
Because of the $aY$ energy splitting,
the suppression of $(1,\pm 1)$ modes sets in with
$B$ field of just a few cG,
and they become quenched for 0.1 G or higher.
The $m_F = 0$ ``unpolarized" modes are oblivious
to the magnetic field until $X$ becomes appreciable,
i.e. for $B\sim a/2\mu_B \sim$ 1kG,
and  get quenched by fields of 1 Tesla or higher.
The scale difference for Fig. 4
would be discussed shortly.
The suppression in $P(\bar M)$
has been taken into account in the experimental limit of eq. (2)
for the interaction of eq. (1).

We have checked and confirmed
the result for the $(V-A)(V+A)$ case \cite{Sasaki},
\begin{eqnarray}
P^{(1,\pm1)}(\bar M) &=& {\bar\delta^2 \over 6(\bar\delta^2 + a^2Y^2 +
\lambda^2)}, \cr
P^{(1,\ \ 0)}(\bar M),\ P^{(0,\ \ 0)}(\bar M)
                                   &=& {\left(2\mp {1\over
\sqrt{1+X^2}}\right)^2 \, \bar\delta^2
                                       \over 6\left[\left(2\mp {1\over
\sqrt{1+X^2}}\right)^2
                                                                \,
\bar\delta^2\, + \lambda^2\right]}.
\end{eqnarray}
The results are also plotted in Figs. 1--4
as ``{\scriptsize $\times$}" symbols.
Conversion in $(0,0)$ mode is
the most prominent \cite{Sasaki},
but gets suppressed by up to a factor of $4/9$ when
the magnetic field goes beyond $\sim$  1kG.
The $(1,\pm1)$ modes are quenched by
magnetic fields of  0.1G or higher,
just like in the $(V-A)(V-A)$ case,
but  $P^{(1,0)}(\bar M)$ actually
{\it grows} with $B$ field around 1kG, and partially
compensates for the drop in $P^{(0,0)}$.

We now state the results for the (pseudo)scalar
induced interaction case.
Details would be given elsewhere \cite{HLW}.
For purely scalar interactions ($f_A = 0$),
we find
\begin{eqnarray}
P^{(1,\pm1)}(\bar M) &=&
                          {\bar\delta^2 \over 2(\bar\delta^2 + a^2Y^2 +
\lambda^2)}, \cr
P^{(1,\ \ 0)}(\bar M)      =    P^{(0,\ \ 0)}(\bar M)    &=&
                          {\bar\delta^2 \over 2(\bar\delta^2  + \lambda^2)}.
\end{eqnarray}
Thus, aside from the familiar quenching of the $(1,\pm1)$ states,
the $(1,0)$ and $(0,0)$ states are
{\it completely insensitive} to magnetic fields \cite{Matthias2}.
For purely pseudoscalar interactions ($f_H = 0$), we find
\begin{eqnarray}
P^{(1,\pm1)}(\bar M) &=& 0, \cr
P^{(1,\ \ 0)}(\bar M),\ P^{(0,\ \ 0)}(\bar M)
                                   &=& {\left(\mp 1 + {1\over
\sqrt{1+X^2}}\right)^2 \, \bar\delta^2
                                       \over 2\left[\left(\mp 1 + {1\over
\sqrt{1+X^2}}\right)^2
                                                                \,
\bar\delta^2\, + \lambda^2\right]}.
\end{eqnarray}
In this case,
muonium conversion occurs solely in the $(0,0)$ mode
for zero magnetic field \cite{HW}.
Conversion in the $(1,0)$ mode starts to grow
from zero for field strengths beyond $\sim$ 1kG,
and partially compensates for
the drop, by a factor of 4, in transition probability in the $(0,0)$ mode.
We plot the results again in Figs. 1--4,
with solid and dashed lines representing
scalar and pseudoscalar case, respectively.

The results for $(S-P)(S-P)$ and $(S-P)(S+P)$ operators
can be similarly obtained.
With the same normalization conditions as described above,
the results are also given in Figs. 1--4 for sake of comparison,
with open circles representing the $(S-P)(S+P)$case  and
open boxes representing the $(S-P)(S-P)$ case.
Note that according to eq. (4),
the $(S-P)(S-P)$ operators should be
subdominant compared to $(S-P)(S+P)$ operators.
Pure $(S\mp P)(S\pm P)$ operators
corresponds to complex neutral scalars \cite{HW},
where the sneutrino $\tilde \nu_\tau$ in
SUSY models with $R$-parity breaking \cite{HM}
as a special case.
It is evident from Figs. 1--4 that the combinations
$(S-P)(S-P)$ and $(S-P)(S+P)$
behave in the same way as
$(V-A)(V-A)$ and $(V-A)(V+A)$, respectively.
In the latter case, the two operators are related to each other
by a Fierz transform.
For the former case,
although the operators can not be related to each other
by a Fierz transform,
the matrix elements are always in same proportion,
which comes as a consequence of the nonrelativistic
limit.

Turning to discussions,
we note that
the assumption of equally populated Breit-Rabi levels is not a valid one,
since this is determined by the muonium formation process
and the magnetic field strength.
However, as a consequence of this 
assumption,
the results in eqs. (19-22) and hence Figs. 2--4
all have an artificial factor of 4.
To illustrate the effect of differently populated Breit-Rabi levels,
normalizing again to the $(V-A)(V-A)$ case at zero $B$ field,
we take muonium states to be populated as \cite{PSI}
32\%, 35\%, 18\% and 15\%, respectively, for
$(F,\ m_F) = (0,\ 0),\ (1, +1),\ (1,\ 0),\ (1,\ -1)$,
and plot the results for $P_T(\bar M)$
(i.e. analogous to Fig. 1) in Fig. 5.
The $(V-A)(V-A)$ and $(S-P)(S-P)$ results are unchanged,
since the conversion matrix elements are the same for
all modes.
The purely scalar ($SS$) case is also unchanged,
since $\vert c_{10}\vert^2 + \vert c_{00}\vert^2
= \vert c_{1+1}\vert^2+ \vert c_{1-1}\vert^2 = 50\%$
is the same as the equally populated case.
The $(V-A)(V+A)$, $(S-P)(S+P)$
and  $PP$ cases
are somewhat modified from the equally populated case,
but the difference for 1kG field is rather slight.

Let us summarize our findings.
Purely scalar ($SS$) induced $M$-$\bar M$ transitions
in polarized (i.e. $(F,m_F) = (1,\pm 1)$) modes are quenched
by magnetic fields of  0.1 Gauss or higher,
but conversion in unpolarized states (i.e. $(1,0)$ and $(0,0)$)
are independent of magnetic field strength.
Purely pseudoscalar ($PP$) interactions
do not induce conversion in $(1,\pm 1)$ states,
but exhibit compensating effects in the
$(1,0)$ and $(0,0)$ channels,
similar to the $(V-A)(V+A)$ case.
Interactions of $(S\mp P)(S\pm P)$ and $(V\mp A)(V\pm A)$ form
have the same magnetic field dependence
since they are related by Fierz transform,
while $(S\mp P)(S\mp P)$ and $(V\mp A)(V\mp A)$ interactions
have the same field dependence
because of
proportional conversion matrix elements.

\acknowledgments
We thank K. Jungmann for numerous communications
and a copy of ref. \cite{Matthias2}.
The work of WSH is supported by grant NSC 84-2112-M-002-011,
and GGW by NSC 84-2811-M-002-035
of the Republic of China.

\vskip -1cm
\figure{
             Magnetic field dependence of total muonium conversion
             probability $P_T(\bar M)$ assuming
              $\vert c_{1,  1}\vert^2
             = \vert c_{1,  0}\vert^2
             = \vert c_{1,-1}\vert^2
             = \vert c_{0,  0}\vert^2
             = 1/4$, and
              normalized to conversion strength of $(V-A)(V-A)$ interaction
              at zero magnetic field.
              Solid and dashed lines stand for $SS$ and $PP$ operators,
              respectively,
              while $\circ$, {\scriptsize $\Box$}, {\scriptsize $+$} and
              {\scriptsize $\times$}
              stand for $(S-P)(S+P)$, $(S-P)(S-P)$,
              $(V-A)(V+A)$ and $(V-A)(V-A)$ cases, respectively.
}
\vskip -1cm
\figure{
             $P^{(1,\pm1)}(\bar M)$ vs. magnetic field with same assumption
             as Fig. 1.
}
\vskip -1cm
\figure{
              $P^{(1,0)}(\bar M)$ vs. magnetic field with same assumption
             as Fig. 1.
}
\vskip -1cm
\figure{
             $P^{(0,0)}(\bar M)$ vs. magnetic field with same assumption
             as Fig. 1.
}
\vskip -1cm
\figure{
             The same as Fig. 1 except
              $\vert c_{1, 1}\vert^2 = 0.35$,
              $\vert c_{1, 0}\vert^2 = 0.18$,
              $\vert c_{1,-1}\vert^2 = 0.15$ and
              $\vert c_{00}\vert^2 = 0.32$.
}

\eject
\end{document}